\documentstyle[12pt]{article}
\newcommand{\dbs}{\renewcommand{\baselinestretch}{2.0}
\large\normalsize}
\begin{document}
\normalsize
\dbs
\setcounter{page}{1}
\title{
Use of the Generalized Gradient
Approximation in Pseudopotential Calculations of Solids
}
\author{Yu-Min Juan and Efthimios Kaxiras\\
{\it Department of Physics and Division of Applied Sciences}\\
{\it Harvard University, Cambridge, MA 02138}\\
Roy G. Gordon\\
{\it Department of Chemistry}\\
{\it Harvard University, Cambridge, MA 02138}\\}
\maketitle
\begin{abstract}
We present a study of the equilibrium properties
of $sp$-bonded solids
within the pseudopotential approach,
employing recently proposed generalized gradient approximation (GGA)
exchange correlation functionals.
We analyze the effects of the gradient corrections on the
behavior of the pseudopotentials
and discuss
possible approaches for constructing
pseudopotentials self-consistently in the context of gradient corrected
functionals.
The calculated equilibrium properties of solids using the GGA functionals are
compared to the ones obtained through
the local density approximation (LDA) and to experimental data.
A significant improvement over the LDA results is achieved
with the use of the GGA functionals for cohesive energies.
For the lattice constant,
the same accuracy as in LDA can be obtained when the nonlinear coupling
between core and valence electrons introduced by the exchange
correlation functionals is properly taken into account.
However, GGA functionals  give bulk moduli that are too small
compared to experiment.
\end{abstract}



There have been extensive efforts recently to improve
the accuracy of density functional theory (DFT) [1]
by going beyond the local density approximation (LDA) [2]. Including
gradient corrections to LDA represents a promising scheme
that is conceptually simple [3]-[10].
This approach is referred to as Generalized Gradient
Approximation, or GGA.
Calculations using different gradient-corrected
functionals have been performed to test the applicability
of this approach on a variety of systems [11]-[17].

In an earlier work [17], we investigated one GGA functional recently proposed
by Perdew and Wang (PW91) [6], [9] in calculations of both atoms and solids.
We showed that
by simply combining the PW91 functional with the pseudopotential approach
leads to lattice constants
for solids, such as simple metals and semiconductors, that
are larger than experiment,
and the percentage errors are significantly larger
than those obtained from LDA.
Results from all-electron calculations [13], [16]
for systems other than the ones we considered, such as transition metals,
suggest that
there is no such significant increase of the error in lattice constants
when the PW91 functional is used.
In order to understand this difference
between the results obtained from the pseudopotential and the
all-electron calculations, we first examined the charge distribution
of individual atoms.
We considered both PW91 and the more recently proposed functional
by Lacks and Gordon (LG) [10].
The reasons for choosing these two functionals are:
(1) They are among the most recent additions to
the list of proposed gradient corrections and are intended
to give better results compared to earlier attempts.
(2) The two functionals have  similar expressions
which simplifies computational implementation.
(3) The Lacks-Gordon exchange functional is  produced by
fitting to exact results, while the PW91 functional is
derived from first principles; comparison between
these two functionals may provide insight for
a more accurate approach.

In studying the charge distributions of individual atoms,
we have found it instructive to evaluate the following quantity
\begin{equation}
C_{nl}(R)=4 \pi \int_{0}^{R} r^{2}dr
\left[\left|\phi^{LDA}_{nl}(r)\right|^{2}-\left|\phi^{GGA}_{nl}(r)\right|^{2}\right]
\end{equation}
$C_{nl}(R)$ is the difference between the charge enclosed
within a sphere of radius $R$ around the nucleus
calculated from the LDA
and the GGA functional for each single electron
orbital $nl$.
In Fig.\ 1 we display $C_{nl}(R)$ for
the $nl=2p$, $3s$ and $3p$ orbitals
of Si, and the $nl=2p$, $3s$, $3p$ orbitals of Na.
The more positive $C_{nl}(R)$ is the more charge has been pushed outside
the region contained by the sphere of radius $R$
in the GGA calculation
compared to
the charge obtained by LDA.

In order to see the difference in the physically important
range, we used the experimentally measured bond length as
the unit along the x-axis in Fig.\ 1.
As is obvious from this figure, there is almost no
difference in the charge distribution between LDA and GGA results
for the $2p$ core orbital of Si.
For the $3s$ and $3p$ valence orbitals of Si,
in the neighborhood of the bonding region
substantial charge has been pushed away from the nucleus
in both the PW91 and
the LG calculations, relative to the LDA results.
A similar situation occurs in the case of Na as is shown in Fig.\ 1.

These comparisons indicate a weaker
interaction between the valence electrons of the atom
and the ion when using GGA functionals as opposed to LDA,
due to the spreading out of the valence charge in GGA calculations.
In an approach which simply replaces the
effects due to the ion and the core electrons with a pseudo-core
which is constructed to reproduce the results of the all-electron calculation,
this will unavoidably lead to a weaker interaction between the valence
electrons
and the pseudo-core.
Since in the pseudopotential framework the properties of solids are
determined by the interaction between
valence electrons and the pseudo-core,
the tendency for valence charge
to be pushed away from the nucleus when GGA
functionals are used leads to a softer solid,
characterized by larger equilibrium
lattice constant and smaller bulk modulus  than LDA.

This observation points to
the necessity of properly
taking into account the nonlinear coupling
between the valence
and core electrons in the exchange correlation functional
within the pseudopotential approach,
when GGA functionals are used.
To examine the effects due to the inclusion of
gradient corrections, we first consider the behavior of the quantity
$s$ which is used
in the definition of GGA functionals
in addition to the charge density $n$.
$s$ is defined as a function of the charge density $n$
and its gradient $\nabla n$:
\begin{equation}
s=\left| \nabla n \right| /2 (3 \pi^{2})^{1/3}n^{4/3}
\end{equation}
Although the charge density $n$ can be separated
into the core charge density $n_{c}$ and the valence charge density $n_{v}$
so that $n=n_{c}+n_{v}$, such a separation can not be written
for $s$
due to
the nonlinearity of the expression of Eq.\ (2).
We use Si as an example to
further illustrate this point.
We display in Fig.\ 2 both the core and valence charge density
for the Si atom,
calculated with the LDA and the GGA functionals respectively.
There is evidently no significant difference between the results of
these two calculations
as far as the overlap between
charge density is concerned.
Since the results from the PW91 and LG functionals tend to be very similar
(see for example the comparison in Fig.\ 1),
from now on we will only present results obtained with PW91.
Fig.\ 3(a) displays the quantity
\begin{equation}
\Delta V_{xc}=V_{xc}[n_{c}]+V_{xc}[n_{v}]-V_{xc}[(n_{c}+n_{v})]
\end{equation}
which is a measure of the nonlinearity of the exchange-correlation potential.
There is a substantial increase in the
nonlinearity of the exchange-correlation potential in the region
where the overlap between the core and valence charge is not negligible
(compare with Fig.\ 2).
This increase in nonlinearity is due to the fact that
the variables used in the expression for the GGA
functionals are not separable in terms of the valence and core
parts. For example, we show in Fig.\ 3(b) the values of
$s_{c}$, $s_{v}$ and $s$ which correspond to the value
of the quantity defined in Eq.\ (2) calculated with the core,
valence, and total charge density respectively.
Therefore, the inclusion of gradient corrections seems to
increase the effects of coupling between the valence and core electrons.

As was discussed in the previous paragraphs, the simple unscreening
of the pseudopotential by
\begin{equation}
V^{ps}_{ion}=V^{ps}_{screened}-V_{H}[n_{v}]-V_{xc}[n_{v}]
\end{equation}
where $V_{H}$ and $V_{xc}$ are the Hartree and
the exchange-correlation potentials respectively,
does not give satisfactory results for the properties of solids.
This approximation implicitly
assumes the linearization of the exchange-correlation functional.
As was pointed out by Louie el al.\ [18], a more consistent approach is
to include the core charge density in the unscreening.
That is,
instead of taking out only the $V_{xc}[n_{v}]$ part in the unscreening
procedure as in Eq.\ (4),
$V^{ps}_{ion}$ should be defined instead as:
\begin{equation}
V^{ps}_{ion}=V^{ps}_{screened}-V_{H}[n_{v}]-V_{xc}[(n_{c}+n_{v})]
\end{equation}
where $n_{c}$ is
a rigid core charge density, constructed from a reference atomic system.
Since the core electrons are not included in
the pseudopotential calculation, whenever the exchange-correlation
energy and potential are needed,
the full charge density $n=n_{c}+n_{v}$ must be used.
This procedure is exact within the rigid core approximation,
but it would require a very large number of plane waves
to describe
the core charge density accurately,
and one loses the advantages of using the pseudopotential formalism by
adopting this approach. So even though
it is theoretically correct, it is not practical from
a computational point of view. Therefore, it is necessary to make some
approximation in order to obtain a practical computational scheme.

In the present paper, we follow the partial core prescription
proposed in Ref.\ [18]. The full core charge density is replaced with an
artificial core charge density $\tilde{n}_{c}$ defined as:
\begin{eqnarray}
\tilde{n}_{c} &=& \frac {A \sin Br} {r}, r \leq  r_{c} \nonumber \\
\tilde{n}_{c} &=& n_{c}, r> r_{c}
\end{eqnarray}
The parameters A and B are
determined by the requirement that the value of $\tilde{n}_{c}$ and its
derivative with respect to the radius $r$ be exactly the same as those of
the real core
charge density $n_{c}$ at the cutoff radius $r_{c}$.
We have found that in order to
capture the nonlinear coupling between the core and the valence electrons
for the case of the PW91 functional,
it is necessary to use $r_{c}$ smaller than what was suggested in Ref.\ [18]
for LDA calculations.
In our calculations, $r_{c}$ is chosen as the radius where the core charge
density $n_{c}$ is 6-7 times larger than the valence
charge density $n_{v}$. In Ref.\ [18] (which dealt
with LDA calculations), $r_{c}$ was chosen as the radius
where the core charge density $n_{c}$ was 2-3 times larger than the
valence charge density.
It is worthwhile mentioning that
the pathological oscillatory behavior of the PW91 exchange-correlation
potential
near the nuclei, which causes problems
in creating smooth pseudopotentials during the
unscreening procedure
[13], [17],
is automatically eliminated by using the
partial core correction.

As an illustration of how this approach works,
we consider four $sp$-bonded solids in their ground state
phase: Si (diamond), Ge (diamond), GaAs (zincblende), and Al (fcc).
We construct self-consistent pseudopotentials
as described in Ref.\ [17] with the partial core prescription
for the unscreening procedure discussed above.
For the LDA calculations we use the exchange-correlation potential
of Ceperley and Alder as parametrized by Perdew and Zunger [19]
and norm-conserving pseudopotentials from Bachelet, Hamann, and
Schl\"{u}ter (BHS) [20].
We use a plane-wave basis for the expansion
of the wavefunction of valence electrons: the highest kinetic energy of
the plane waves in the basis is
16 Ry.
For reciprocal space integrations,
29 special k-points in the irreducible Brillouin zone are used for the diamond
structure and the zinc-blende structure,
and 213 special k-points for the fcc
structure.
The gradient and the laplacian of the density, which are needed
for the GGA functionals considered here, were obtained through
FFT's,
with minimal increase in CPU time
(less than 3\%).
The calculated energy vs.\ volume results are fitted to the Birch-Murnaghan
equation of state [21].
The equilibrium properties are then derived from
the equation of state curves.
The cohesive energy is taken to be the total energy difference between
the solid in equilibrium and the isolated atom.
Spin polarization
effects on the free atom energy
are taken into account with the empirical formula $\Delta E_{p}=-0.18 \times
n_{p}^{2}$, [22], [23], where $n_{p}=n_{\uparrow}-n_{\downarrow}$ with
$n_{\uparrow}$
the number of electrons having spin up and $n_{\downarrow}$ the number of
electrons with spin down. Spin polarization is
expected to have negligible effects on the total energy $E_{0}$ of
nonmagnetic solids.

The calculated ground state properties
using LDA and PW91
are summarized and compared to experimental data
in Table I.
The results from PW91
represent a substantial improvement of the over-binding
problem of LDA: the cohesive energies are in better agreement
with experiment for all the solids we have considered.
This improvement in cohesive energies
can be attributed to a large extene to the fact PW91 gives a more accurate
atomic
energy than LDA [17].
For the equilibrium lattice constant, the value obtained from PW91
is consistently larger than LDA results.
In the case of Al, this makes the value obtained from PW91 closer
to experiment than the LDA result.
For Si, Ge, and GaAs,
the results obtained from LDA and PW91 are
of the same accuracy compared to experiment.
PW91 tends to overcorrect the LDA results and gives an overestimate
for the equilibrium lattice constant of these systems.
For the bulk modulus, the values obtained from PW91 are smaller
than the  LDA results.
While this leads to a better result for Al, the bulk
moduli we obtained for the three semiconductors are
significantly underestimated (by -12 \% to -25 \% compared to experiment).
Similar
observations for earlier gradient-corrected functionals have
been reported [11], [12].
Finally, we compare our results to recent all-electron, linearized augamented
planewave (LAPW) [24] total energy and electronic structure calculations, with
the same GGA functional as in our work.
It is obvious from the comparison of Table I that the present
pseudopotential calculation results with the partial core correction
represent a significant improvement over
the results without the partial core correction,
and agree very well with the all-electron LAPW calculation results.
The remaining discrepancy is probably
due to the relaxation of the core electrons which is not allowed
in the pseudopotential calculation.

For the electronic structure, we compare in Table II the band gap predicted
by LDA and PW91 at
both the experimentally measured lattice constant and  the theoretical
equilibrium lattice constant.
An overall improvement which brings the
values closer to the experimental data at the experimental
lattice constant is found
for all three semiconductors we have
considered, although the magnitude of the improvement depends
on the material. There is no consistent improvement for the
band gaps at the theoretical equilibrium lattice constant.
We note here that this is simply
a comparison between LDA and PW91 as different approximations
to the exchange correlation functionals. The well known inability
of density functional theory to reproduce accurately band gaps in
semiconductors
and insulators is much more complicated
and  related to the intrinsic discontinuity of the exchange correlation
functional [25], which is not represented by either of the two approximations
used here.
Good agreement
for the band gap values nevertheless can be obtained by using DFT/LDA
wavefunctions and
solving the self-energy operator equations, within the so called
GW approximation [26].

In conclusion,
we showed that it is essential to take into account
the core-valence coupling in the pseudopotential
calculations when using GGA exchange-correlation functionals.
To this end, we have found that the partial core prescription of Louie et al.\
[18] is most appropriate
when using a plane-wave basis.
We considered the structural properties
of Si, Ge, GaAs, and Al
using both LDA and PW91.
We found that PW91
gives consistently better cohesive energies than LDA.
We also demonstrated that for the lattice
constant the same accuracy as in LDA can
be obtained with GGA, as long as the nonlinearity
of the gradient corrected functional is properly
taken into account.
For the semiconductors we considered, the bulk moduli obtained
with the use of GGA functionals
represent significant underestimates of the experimental results.
The PW91 functional does give a better description for the equilibrium
properties of Al.
We conclude that further search may be needed
for an exchange-correlation functional which is consistently
better
for all solids.

In view of the above results, one may inquire what are the physical
situations in which the use of GGA functionals can provide significant
improvements over LDA results. Recently,
calculations
have been reported for $H_{2}$ dissociation on a Cu(111) surface with the LDA
and the GGA [27], [28]. The GGA results for this system represent
significant improvements over
the LDA results.
It has also been demonstrated that the GGA gives results
in better agreement with experiments than the LDA for finite systems (atoms
and molecules) and metallic surfaces [13]-[15].
It is therefore expected that the GGA will give, in general,
a better description
for the interaction between molecules and other molecules or solid surfaces.
The reason that the GGA should give better results for these interactions
can be attributed to the fact that
substantial part of the interactions in
the these systems are related to the tails of the electronic wave functions,
where
the GGA gives a more accurate description than the LDA.

This work was supported by the Materials Research Laboratory
of Harvard University
which is funded by National Science Foundation Grant \# DMR 89-20490.
The calculations were carried out at the Cornell National Supercomputer
Facility.

\section*{References}

\hspace{0.265in}[1]
P. Hohenberg and W. Kohn,
{\em Phys. Rev.} {\bf 136}, B864 (1964).

[2]
W. Kohn and L. Sham,
{\em Phys. Rev.} {\bf 140}, A1133 (1965).

[3]
D.C. Langreth and M.J. Mehl,
{\em Phys. Rev.} {\bf 28}, 1809 (1983);
D.C. Langreth, in
{\em Many-Body Phenomena at Surfaces,} edited by D.C. Langreth and H. Suld
(Plenum, New York, 1984).

[4]
{\em Theory of the Inhomogeneous Electron Gas,}
edited by S. Lundqvist and N. H. March
(Plenum, New York, 1983).

[5]
C.D. Hu and D.C. Langreth,
{\em Phys. Scripta.} {\bf 32}, 391 (1985).

[6]
J.P. Perdew,
{\em Phys. Rev. Lett.} {\bf 55}, 1665 (1985);
J.P. Perdew and Y. Wang,
{\em Phys. Rev. B} {\bf 33}, 8800 (1986).

[7]
A.D. Becke,
{\em Phys. Rev. A} {\bf 38}, 3098 (1988);
F.W. Kutzler and G.S. Painter,
{\em Phys. Rev. Lett.} {\bf 59}, 1285 (1987).

[8]
M. Norman and D.D. Koelling,
{\em Phys. Rev. B} {\bf 28}, 4357 (1983).

[9]
J.P. Perdew,
in {\em Electronic
Structure of Solids '91,} edited by P. Ziesche and H. Eschrig
(Akademie Verlag, Berlin, 1991).

[10]
D.J. Lacks, and R.G. Gordon,
{\em Phys. Rev. A} {\bf 47}, 4861 (1993).

[11]
B. Barbiellini, E.G. Moroni, and T. Jarlborg,
{\em J. Phys.: Condens. Matter} {\bf 2}, 7597 (1990).

[12]
A. Garc\'{i}a, C. El\H{a}sser, J. Zhu, S. Louie, and M.L. Cohen,
{\em Phys. Rev. B} {\bf 46}, 9829 (1992).

[13]
G. Ortiz and P. Ballone,
{\em Phys. Rev. B} {\bf 43}, 6376 (1991).

[14]
J.P. Perdew, J.A. Chevary, S.H. Vosko, K.A. Jackson, M.R. Pederson,
D.J. Singh, and C. Fiolhais
{\em Phys. Rev. B} {\bf 46}, 6671 (1992).

[15]
B.G. Johnson, P.M.W. Gill, and J.A. Pople,
{\em J. Chem. Phys.} {\bf 98} (7), 5612 (1993).

[16]
V. Ozolins and M. Korling,
{\em Phys. Rev. B} {\bf 48}, 18304 (1993).

[17]
Y.M. Juan and E. Kaxiras,
{\em Phys. Rev. B} {\bf 48}, 14944 (1993).

[18]
S.G. Louie, S. Froyen, and M.L. Cohen,
{\em Phys. Rev. B} {\bf 26}, 1738 (1982).

[19]
D.M. Ceperley and B.J. Alder,
{\em Phys. Rev. Lett.} {\bf 45}, 566 (1980);
J. Perdew  and A. Zunger,
{\em Phys. Rev. B} {\bf 23}, 5048 (1981).

[20]
G.B. Bachelet, D.R. Hamann, and Schl\"{u}ter,
{\em Phys. Rev. B} {\bf 26}, 4199 (1982).

[21]
F.D. Murnaghan,
{\em Proc. Nat. Acad. Sci.} {\bf 30}, 244 (1944).

[22]
O. Gunnarson, B.I. Lundqvist, and J.W. Wilkins,
{\em Phys. Rev. B} {\bf 10}, 1319 (1974).

[23]
J. Ihm and J.D. Joannopoulos,
{\em Phys. Rev. B} {\bf 24}, 4191 (1981).

[24]
C. Filippi, D.J. Singh, and C.J. Umrigar
(to be published in {\em Phys. Rev. B}).

[25]
J. Perdew and M. Levy, {\em Phys. Rev. Lett.} {\bf 51}, 1884 (1983);
L. Sham and M. Schl\"{u}ter, {\em Phys. Rev. Latt.} {\bf 51}, 1888 (1983);
M. Lannoo, M. Schl\"{u}ter and L.J. Sham, {\em Phys. Rev. B} {\bf 32}, 3890
(1985).

[26]
L. Hedin, {\em Phys. Rev.} {\bf 139}, A796 (1965);
M. Hybertsen and S. Louie, {\em Phys. Rev. Lett} {\bf 55}  1418 (1985);
M. Hybertsen and S. Louie, {\em Phys. Rev. B} {\bf 34}  5390 (1986);
R. Godby, M. Schl\"{u}ter, and L. Sham, {\em Phys. Rev. Lett.} {\bf 56}, 2415,
(1986);
R. Godby, M. Schl\"{u}ter, and L. Sham, {\em Phys. Rev. B} {\bf 37}, 10159,
(1988).

[27]
B. Hammer, M. Scheffler, K.W. Jacobsen, And J.K. N{\o}rskov,
{\em Phys. Rev. Lett.} {\bf 73}, 1400 (1994).

[28]
J.A. White, D.M. Bird, M.C. Payne, and I. Stich,
{\em Phys. Rev. Lett.} {\bf 73}, 1404 (1994).

\section*{Figure Captions}
\begin{enumerate}

\item{
Integrated charge difference $C_{nl}(R)$ [see Eq.\ (1)] calculated with PW91
(solid lines),
and LG (dashed lines) for
$nl=$$2p$, $3s$, $3p$ orbitals of Si, and
$nl=$$2p$, $3s$, $3p$ orbitals of Na. The results for the $3p$ orbital of
Na
have been divided by a factor 2 so that they can be displayed on the same scale
as the results for Si}

\item{
The core (upper panel) and valence (lower panel)
charge density of the silicon atom calculated with LDA
and GGA functionals as a function of  distance from the nucleus.
Notice the different density scales in the two panels.
}

\item{
(a) The quantity $\Delta V_{xc}$ [see Eq.\ (3)] for the  silicon atom
calculated with LDA and PW91 as a function of
distance from the nucleus.

(b)The values of $s$ [see Eq.\ [2]]
calculated from the core ($s_{c}$), the valence ($s_{v}$),
and the total  charge density ($s$) for the silicon atom as a function
of
distance from the nucleus.
}
\end{enumerate}
\end{document}